\documentclass[10pt,a4paper]{article}
\usepackage{graphicx}
\usepackage{amsmath}

\usepackage[margin=2.5cm]{geometry}

\begin{document}

\setlength{\parindent}{0.0em}

{\sf \LARGE Coupling carbon nanotube mechanics to a superconducting
circuit }

\vspace{0.5em}

B.H. Schneider$^1$, S. Etaki$^1$, H.S.J. van der Zant$^1$, G.A. Steele$^1$

\vspace{1em}

{\em $^1$Kavli Institute of NanoScience, Delft University of
Technology, PO Box 5046, 2600 GA, Delft, The Netherlands.}

\vspace{1em}

{ \bf
   
The quantum behaviour of mechanical resonators is a new and emerging
field driven by recent experiments reaching the quantum ground state.
The high frequency, small mass, and large quality-factor of carbon
nanotube resonators make them attractive for quantum nanomechanical
applications.  A common element in experiments achieving the resonator
ground state is a second quantum system, such as coherent photons or
superconducting device, coupled to the resonators motion.  For
nanotubes, however, this is a challenge due to their small size.
Here, we couple a carbon nanoelectromechanical (NEMS) device to a
superconducting circuit.  Suspended carbon nanotubes act as both
superconducting junctions and moving elements in a Superconducting
Quantum Interference Device (SQUID).  We observe a strong modulation
of the flux through the SQUID from displacements of the nanotube.
Incorporating this SQUID into superconducting resonators and qubits
should enable the detection and manipulation of nanotube mechanical
quantum states at the single-phonon level.
 
}

\vspace{2em}

The remarkable properties of nanoelectromechanical systems (NEMS) are
useful for a wide variety of applications, such as ultra-sensitive
force detection
\cite{rugar2004single,lehnert,usenko2011superconducting}, mass
detection at the single atom level
\cite{jensen2008atomic,chaste2012nanomechanical}, and exploring the
quantum limit of mechanical motion
\cite{schwab2005putting,poot,Cleland,teufel,painter}. For all of these
applications, sensitive detection of the resonator motion is
crucial. The ultimate limit of the detection of the motion of a
mechanical resonator is given by its quantum zero point fluctuations,
which result in an uncertainty in the resonators position determined
by the standard quantum limit\cite{teufel2009nanomechanical}.

\setlength{\parindent}{2em}

An effective way of detecting the quantum motion of mechanical devices
is to couple their displacement to another ``probe'' quantum system which
can be read out and controlled, 
such as coherent quantum states of a superconducting qubit
\cite{Cleland}, coherent photons in a cold microwave resonator circuit
\cite{teufel}, or the coherent fields of a laser \cite{painter}.
In the last decade, superconducting circuits
have emerged as an established platform for engineering and
controlling quantum behaviour \cite{clarke2008superconducting}.  This
has formed the motivation for many works coupling MEMS and NEMS
devices to superconducting circuits
\cite{naik2006cooling,lehnert,teufel2009nanomechanical,rocheleau2009preparation,lahaye2009nanomechanical,Cleland,teufel}.
By coupling microelectromechanical (MEMS) devices to quantum
superconducting circuits, researchers have recently demonstrated
mechanical resonators in their quantum ground state
\cite{Cleland,teufel} and achieved single-phonon control over their
motion \cite{Cleland}.  An outstanding challenge in these experiments
is simultaneously achieving both a strong coupling at the
single-phonon level, together with a high frequency and large quality
factor for the mechanical resonator.

Carbon nanotube mechanical resonators
\cite{sazonova2004tunable,garcia2007mechanical,huettel2009carbon}
posses a unique combination of large quality factor, small mass, and
high frequency. Because of their small size, however, it is not easy
to couple to their quantum motion. Techniques based on a capacitive
interaction with superconducting qubits or microwave photon cavities,
as was done in recent experiments with MEMS devices reaching the
mechanical quantum ground state \cite{Cleland,teufel}, do not provide
sufficient coupling strength. An alternative is to incorporate the
mechanical element into a SQUID, as was demonstrated recently using
micromechnical beams \cite{Etaki}. If a carbon nanotube NEMS element
could be included in a SQUID, the SQUID could then be used as a
transducer to couple the mechanical motion to a superconducting cavity
\cite{buks2007displacement}, or as the basis for a superconducting
qubit, coupling the motion directly to the quantum states of the
qubit. 

Here, we demonstrate the coupling of a carbon nanotube NEMS device to
a superconducting circuit, based on a suspended carbon nanotube
SQUID. The flux through the SQUID couples to the nanotube displacement
with a strength of 0.36 m$\Phi_0$/pm. This new device opens up the
possibility of combining carbon NEMS devices with the quantum toolkit
from the superconducting community. Doing so, we expect the suspended
nanotube SQUID will provide a platform for detection of the nanotube
resonator's ground state, and control over its motion at the level of
single phonons.

\vspace{2em}
\setlength{\parindent}{0.0em}
{\bf \large  Results}

The device consists of a SQUID in which the two
Superconductor-Normal metal-Superconductor (SNS) weak links are made
from carbon nanotube junctions \cite{Pablo,Wernsdorfer,NingLau}.  In
contrast to earlier works, the carbon nanotubes are freely suspended,
and thus also act as NEMS elements embedded in the SQUID.  To make the
suspended nanotube SQUID, a clean carbon nanotube is grown in the last
step of fabrication  \cite{steele2009tunable} over a trench patterned
between metal source and drain contacts made from a MoRe
superconducting alloy (see Methods).  A Scanning Electron Microscope
(SEM) image of a typical device is shown in Fig.\ 1a.  The doped
substrate below the trench is used as a global backgate.  The device
is mounted in a superconducting magnet coil with the magnetic field
aligned in the plane of the sample, parallel to the trenches, as
indicated in Fig.\ 1b.  The magnetic field orientation is chosen to
maximize the coupling of vertical displacements on the nanotube to
flux in the SQUID loop \cite{Etaki}. A small misalignment of the
sample also induces a magnetic field perpendicular to the
sample surface which is used to tune the flux operating point of the
SQUID.

\begin{figure}[hb]
\begin{center}
\includegraphics[width=\textwidth]{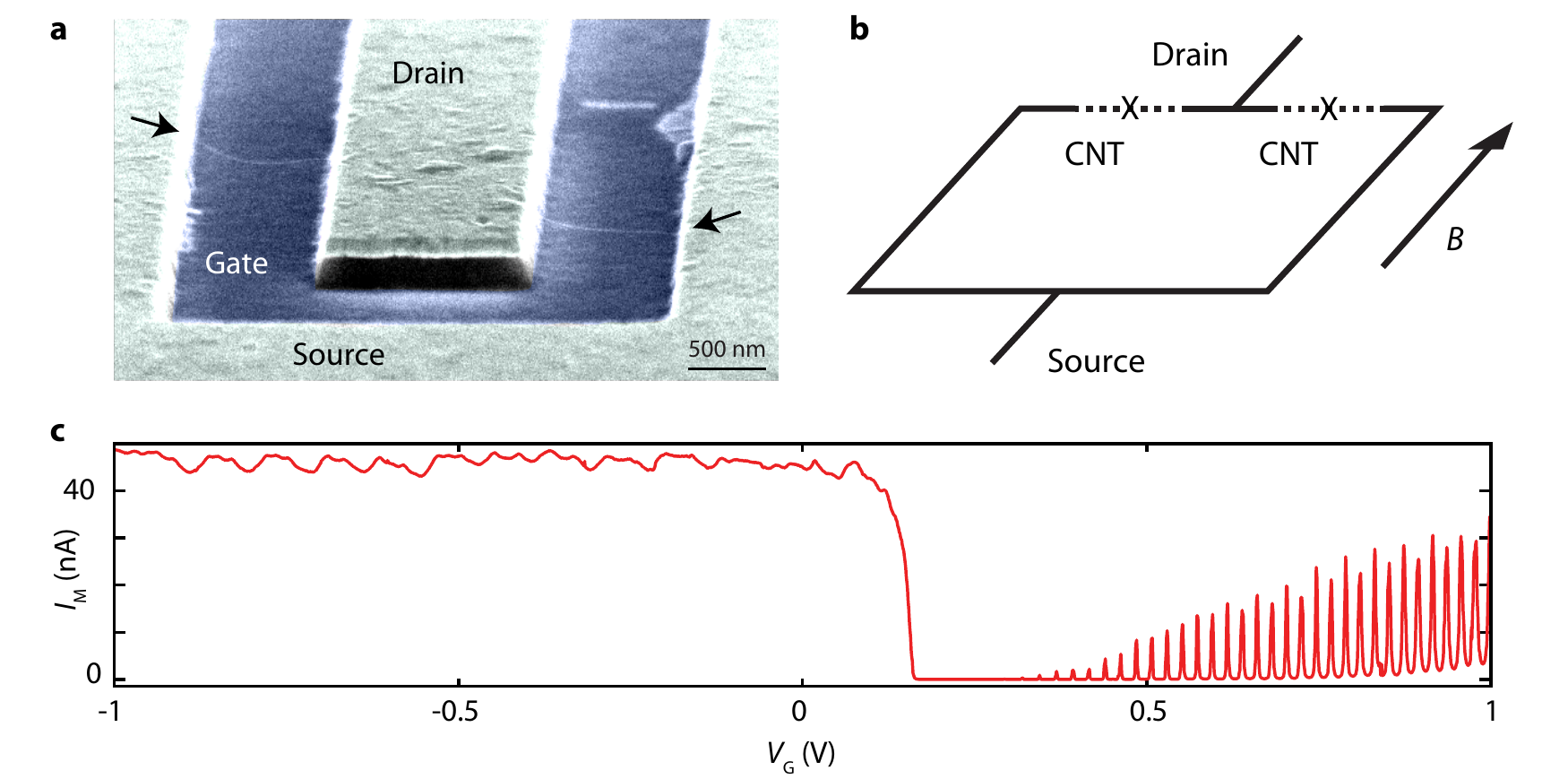}
\end{center}
\caption{ \textbf{Design and characterization of the device a,}
  Colorised scanning electron microscope (SEM) image of a typical
  device showing a single carbon nanotube (CNT) crossing two trenches.
  The device is made by etching two 800 nm wide trenches in a MoRe
  layer and underlying SiO$_{2}$, and subsequently growing CNTs over
  the prefabricated structure.  We apply a voltage $V_{\text{G}}$ to
  the doped substrate underneath the SiO$_2$, which acts as a
  gate. \textbf{b,} Schematic drawing of the SQUID.  The two suspended
  segments of the CNT form two SNS-type Josephson junctions, indicated
  by the crosses. An external magnetic field $B$ can be applied
  in-plane along the trench. \textbf{c,} $I_{\text{M}}$ as a function
  of $V_{\text{G}}$ measured with an applied source-drain bias voltage
  $V_B = 2$ mV.  For $V_{\text{G}} > 0.3$ V, the suspended segments
  form a p-n-p quantum dots exhibiting Coulomb blockade. For
  $V_{\text{G}} < 0.2$ V, the CNT is doped with holes, showing
  Fabry-Perot oscillations with high conductance.}

\label{fig1_main}%
\end{figure}

\setlength{\parindent}{2em}

Figure 1c shows the current through the device ($I_{\text{M}}$) as a
function of the applied gate voltage ($V_{\text{G}}$).  From the gate
voltage thresholds for electron and hole conduction, we estimate that
the carbon nanotube has a small bandgap of 40 meV (see Supplementary
Information).  Due to a work function difference present between the
nanotube and the metal contacts, the nanotube is doped with holes near
the edge of the trench.  At positive gate voltages, electrons induced
by the gate are confined by p-n junction tunnel
barriers \cite{steele2009tunable} in a Coulomb-blockaded quantum dot.
For negative gate voltages, holes are induced in the suspended segment
with no tunnel barriers.  Here, instead of Coulomb blockade,
conductance oscillations arising from Fabry-Perot electronic
interference \cite{bockrath} are observed with conductances of up to
$4.7$ e$^2$/h (see Fig.\ 2a).  Note that this value exceeds the
maximum conductance expected for a single carbon nanotube
($G_\text{max}$ = 4 e$^2$/h), consistent with a SQUID geometry
(Fig.\ 1b) where there are two carbon nanotube junctions in parallel
(Fig.\ 1a).

\begin{figure}[hb]
\begin{center}
\includegraphics[width=\textwidth]{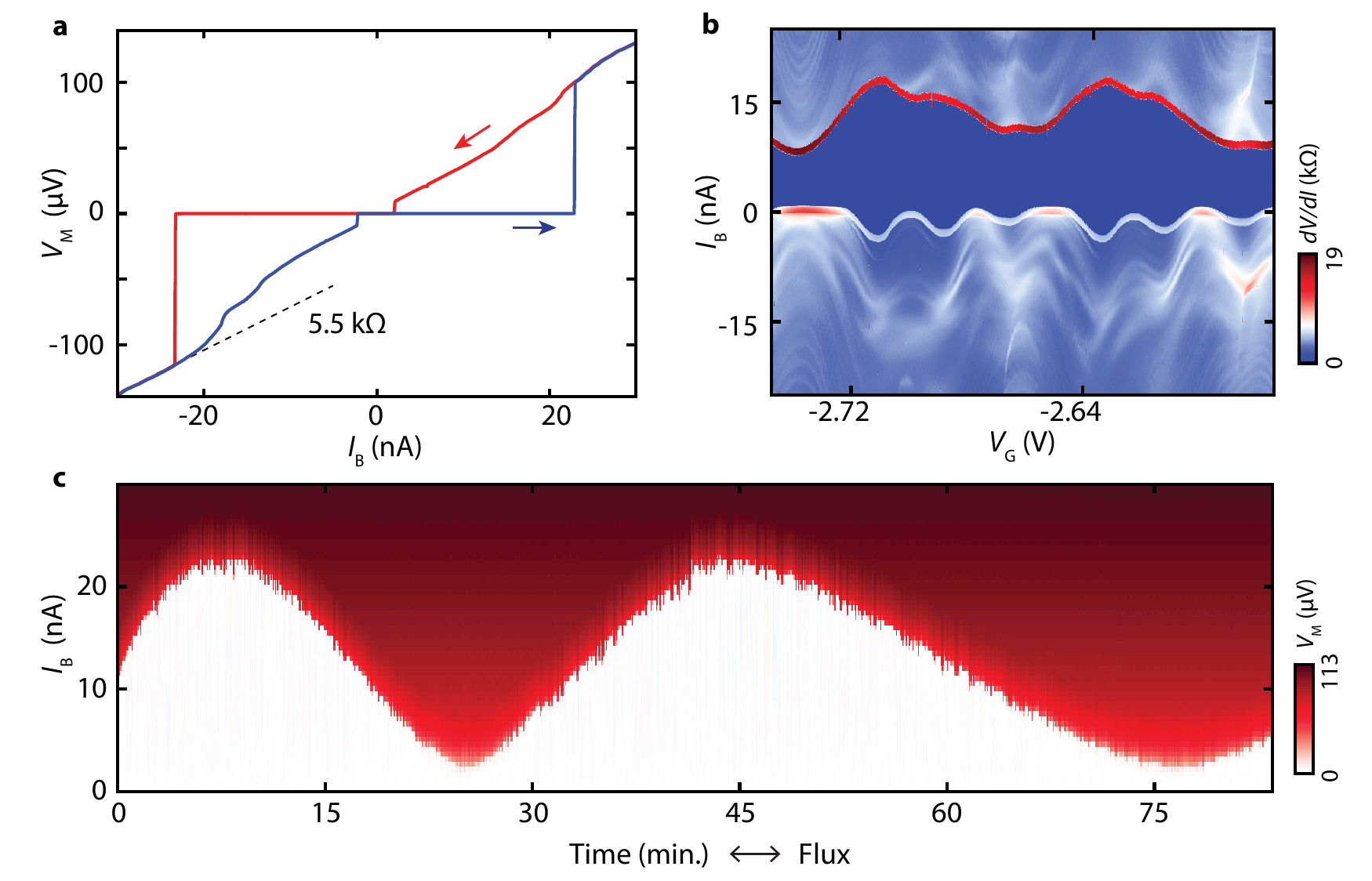}
\end{center}
\caption{\textbf{A suspended carbon nanotube SQUID a,} Four-terminal
  current-voltage trace at $V_{\text{G}} = -1.1$ V.  The onset of the
  resistive branch occurs at 24 nA, corresponding to a critical
  current of 12 nA per nanotube junction.  The dashed line indicates
  the normal-state resistance of 5.5 kOhm ($G = 4.7$ e$^2$/h).  The
  device is hysteretic, as can be seen from the forward (blue) and
  reverse (red) sweep directions  \cite{clarke2006squid}. \textbf{b,}
  Differential resistance $dV_\text{M}/dI_\text{B}$ (color map) as a
  function of $I_{\text{B}}$ and $V_{\text{G}}$ (forward sweep
  direction).  The critical current (red stripe at positive
  $I_{\text{B}}$) oscillates with $V_{\text{G}}$, following the
  modulation of the normal state conductance.  \textbf{c,} SQUID voltage
  $V_{\text{M}}$ as a function of applied current $I_{\text{B}}$ and
  time, taken at $B = 0$ and $V_G = -1.1$ V. Flux creep in the
  superconducting magnet coil is used to apply a small time varying
  magnetic field.  The critical current oscillates from 2 nA up to 24
  nA as a function of the flux through the SQUID.  The near complete
  suppression of $I_{\text{C}}$ at the minima indicates that the SQUID
  has highly symmetric junctions.}
\label{fig2_main}%
\end{figure}

In Fig.\ 2, we use the gate to dope the nanotube with holes such that
the device is tuned into the high-conductance Fabry-Perot regime, and
measure the voltage across the SQUID as a function of an applied
bias current (Fig.\ 2a).  At low currents, the voltage across
the device is zero, a clear indication of a proximity effect induced
supercurrent.  At a critical current of 24 nA, there is a switch to a
finite voltage state.  We attribute the large critical current in our
junctions (12 nA per nanotube) to the high critical temperature of the
superconducting metal leads ($T_{\text{C}} = 5.5$ K), the low contact
resistance between the superconducting metal and the nanotube, and the
clean electronic characteristics of the carbon nanotube.  As shown in
Fig.\ 2b, the critical current is strongly modulated by the gate through
to gate dependence of the normal-state conductance \cite{Pablo}.  Due
to the high transparency of the superconductor-nanotube interface, the
supercurrent in our device also remains finite in the valleys between
the peaks in conductance.

To demonstrate that the device acts as a SQUID, we measure the
critical current as a function of the flux through the loop.  In
practice, we do this by sweeping a large in-plane magnetic field to
zero and then subsequently measure the critical current as a function
of time.  Due to creep in the superconducting magnet coil and the
slight misalignment of the field to the sample plane, there is a small
magnetic field component perpendicular to the surface that continues
to change slowly after the sweep is completed. In such a measurement,
shown in Fig.\ 2c, the observed critical current oscillates between a
value of 24 nA and 2 nA.  These oscillations result from quantum
interference of the superconducting wavefunction traversing the two
junctions of the SQUID  \cite{clarke2006squid}. The large sensitivity
of the critical current to the flux from the small magnetic field
creep, together with the single periodicity, is consistent with a
single large SQUID loop formed by one nanotube SNS junction across
each trench, as shown in the SEM image of a typical device in
Fig.\ 1a.

\begin{figure}
\begin{center}
\includegraphics[width=\textwidth]{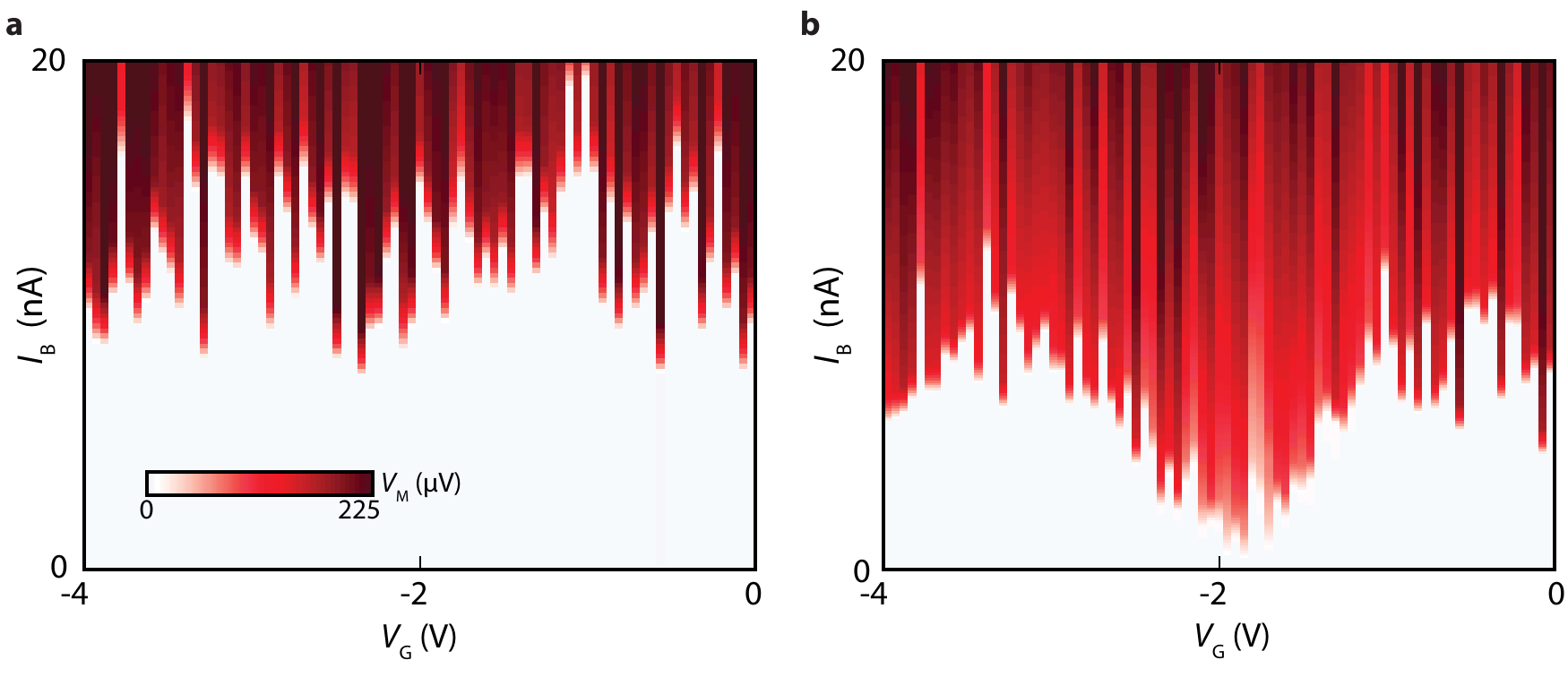}
\end{center}
\caption{ \textbf{A gate-voltage induced magnetic flux a,} 
	Colormap of $V_M$ as a function of $I_{\text{B}}$ and
  $V_{\text{G}}$, taken at $B = 0$ T.  The critical current
  $I_{\text{C}}$ is defined by the onset of a finite voltage state,
  indicated by the red regions. The Fabry-Perot modulation of the
  critical current (Fig.\ 2b) is visible as rapid
  fluctuations in $I_{\text{C}}$ due to the large steps in
  $V_{\text{G}}$.  The measurement is performed fast enough such that the
  magnetic field creep (Fig.\ 2c) can be neglected. \textbf{b,} Same
  measurement taken at $B = 250$ mT. An additional approximately
  sinusoidal modulation of the critical current is seen with a gate
  periodicity of $\Delta V_{\text{G}} = 2.5$ V.}
\label{fig3_main}%
\end{figure}

We now turn our attention to the behaviour of the device in the
presence of a static in-plane magnetic field applied parallel to the
trench. Figures 3a and 3b show the gate voltage dependence of the
critical current with and without an external magnetic field.  The
measurements are taken intentionally with a coarse gate voltage
resolution such that the measurement time is short compared to the
timescale of the flux creep (see Supplementary Information for further
details). The steps in gate voltage are much larger than the
periodicity of the Fabry-Perot conductance oscillations (Fig.\ 2a);
consequently, these appear in Fig.\ 3a as (reproducible) single-pixel
fluctuations.  Figure 3b shows the same measurement taken at an
external magnetic field of 250 mT. Here, an additional nearly
sinusoidal modulation of the critical current can be seen as a
function of gate voltage with a periodicity much longer than that of
the Fabry-Perot conductance oscillations.

In the following, we show how this additional gate modulation of the
critical current arises from a change in magnetic flux induced by the
d.c.\ gate voltage.  The mechanism for such a gate induced flux is
illustrated in Fig.\ 4b.  Increasing the gate voltage, the nanotube is
pulled towards the gate by the attractive electrostatic force between
them.  The nanotube displacement introduces an extra area $\Delta A$
of the SQUID loop perpendicular to the magnetic field aligned along
the trench.  This area change results in a flux change linearly
proportional to the displacement, $\Delta \Phi \propto B \ell u$,
where $B$ is the in-plane magnetic field, $\ell$ is the nanotube
length, and $u$ is the vertical displacement of the nanotube. If the
displacement $u$ is linear in the gate voltage (as expected for
certain gate voltage ranges, see Supplementary Information for further
discussion), the critical current of the SQUID will then oscillate as
a function of gate voltage with constant periodicity, as observed in
Fig.\ 3b.

\begin{figure}[ht]
\begin{center}
\includegraphics[width=\textwidth]{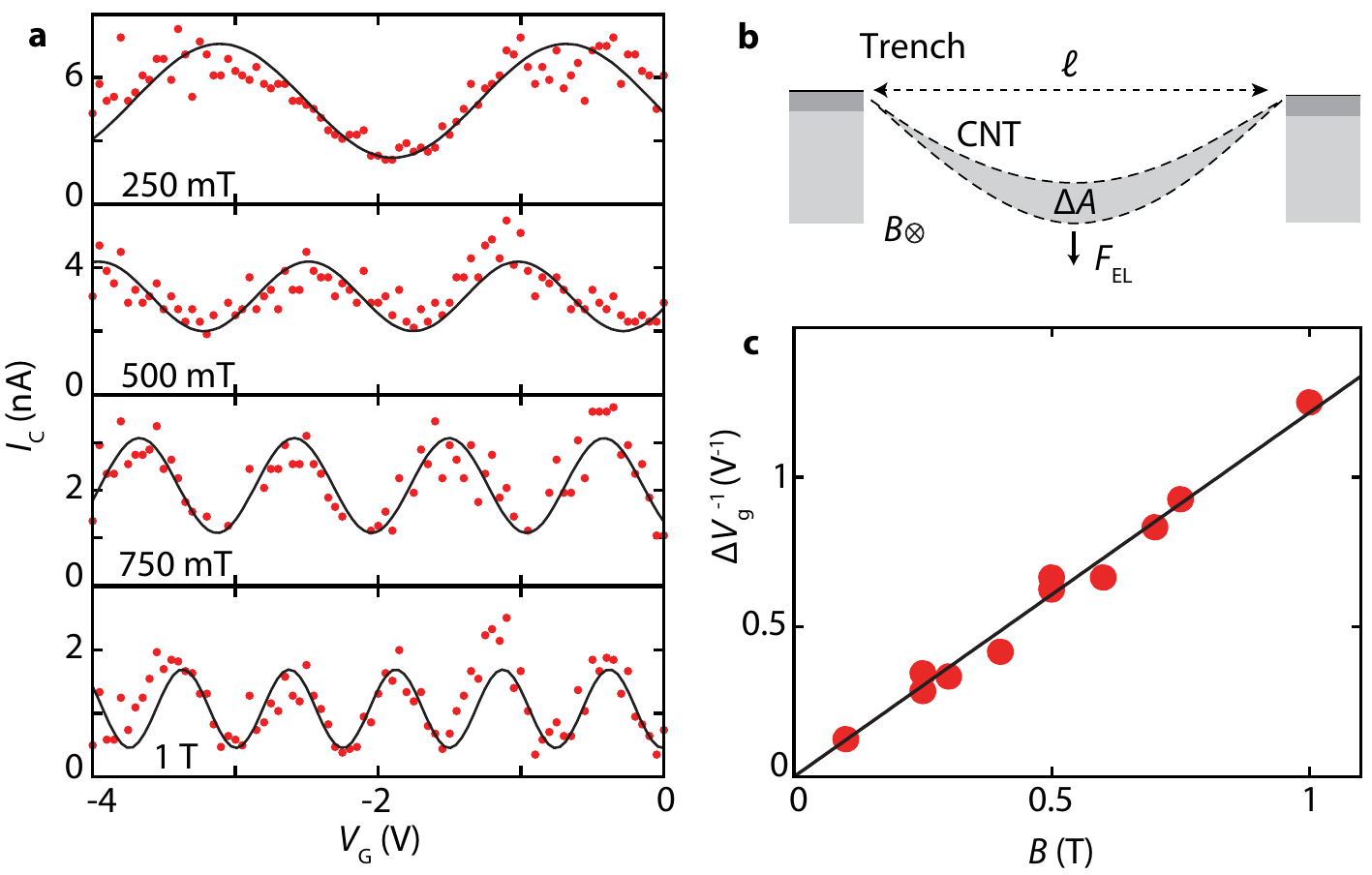}
\end{center}
\caption{ \textbf{Mechanical flux coupling a,} 
	Extracted $I_{\text{C}}$ (dots)
  as a function of $V_{\text{G}}$ at four different magnetic
  fields.  Lines show a sinusoidal fit to the data (red points) from
  which the periodicity $\Delta V_{\text{G}}^{-1}$ is extracted. At $B
  = 1$ T, five oscillations are visible corresponding to a flux
  change of 5 $\Phi_{0}$.  \textbf{b,} Vertical displacement of the nanotube
  due to the electrostatic force from the gate changes the area of the
  SQUID loop perpendicular to the in-plane magnetic field.  This
  produces a gate-induced flux change proportional to the nanotube
  displacement, resulting in an $I_{\text{C}}$ that oscillates 
  with $V_G$.  We estimate that the nanotube displaces by 7.4 nm as
  $V_{\text{G}}$ is swept from 0 to -4 V ($\Delta A \sim 5.5 \times 10^{-3}\ \mu$m$^2$ per
  nanotube). \textbf{c,} $\Delta V_{\text{G}}^{-1}$ (dots) as a
  function of magnetic field with a linear fit (black line).  At $B =
  1$ T, the flux couples to the vertical displacement of the nanotube
  with a flux coupling of $0.36$ m$\Phi_{0}/\text{pm}$.}
\label{fig4_main}%
\end{figure} %

If this gate-voltage induced flux indeed arises from a mechanical
displacement of the nanotube, the flux coupling should scale linearly
with the external magnetic field.  Figure 4a shows the extracted
$I_{\text{C}}$ versus $V_{\text{G}}$ for fields up to 1 T.  To extract
the gate periodicity of the flux oscillation, the critical current as
a function of gate voltage is fit to a sinusoidal function in order to
approximate the expected oscillatory SQUID response
 \cite{clarke2006squid}.  The resulting gate frequency
$V_{\text{G}}^{-1}$ is plotted as a function of magnetic field in
Fig.\ 4c.  The linear scaling of the periodicity with magnetic
field demonstrates that the modulation is due to a flux originating
from the mechanical displacement of the nanotube.  At magnetic fields
of 1 T, we couple the motion of the nanotube to the flux in the SQUID
with a coupling strength of 0.36 m$\Phi_0 $/pm (see Supplementary Information).

\vspace{2em}
\setlength{\parindent}{0.0em}
{\bf \large Discussion}

In the previous section, we demonstrate a strong coupling of the flux
in a SQUID to the displacement of the carbon nanotube NEMS device. The
strong flux coupling is, on its own, not unique to our device: for
example, a larger mechanical flux responsivity was observed in a top-down
fabricated SQUID \cite{Etaki}. What is unique to our device is the
combination of such a flux coupling with a mechanical resonator with
small mass (attogram) and large zero point fluctuations.  This can be
quantified in terms of the amount of flux noise the mechanical
zero-point fluctuations would induce in the SQUID.  The expected zero
point motion of the nanotube is on the order of $u_{\text{ZPF}} =
\sqrt{\hbar/ 2 m \omega} = 3.6$ pm. Together with a quality factor of
$3\times10^4$ and a mechanical resonance frequency of 125 MHz observed
in this device (see Supplementary Information), this results in a peak
in the flux noise spectrum of the SQUID at the mechanical resonance
frequency with an amplitude of 16 $\mu\Phi_{0}/\sqrt{\text{Hz}}$. This
noise level, corresponding to the imprecision noise from the standard
quantum limit for our device, is nearly two orders of magnitude larger
than the 0.2 $\mu\Phi_{0}/\sqrt{\text{Hz}}$ sensitivity that has been
demonstrated coupling a SQUID to a superconducting resonator
\cite{hatridge2011dispersive}. Doing so, we expect that such a
high frequency suspended carbon nanotube SQUID can be used as a linear
position detector with an imprecision noise below the standard quantum
limit, enabling the detection of the quantum motion of a carbon
nanotube.

\setlength{\parindent}{2em}

Finally, the strong coupling between flux and nanotube displacement
could also be used to implement a nanomechanical resonator coupled to
a superconducting qubit.  The important characterization of the
coupling between the two quantum systems is the zero-phonon coupling
rate $g$, given by the energy shift of the probe quantum system in
response to the zero-point fluctuations of the mechanical
device \cite{painter}. In order to have a coherent interaction between
the probe and the mechanical system at the single-phonon level, the
coupling rate $g$ must be larger than the decoherence rates of quantum
states of the mechanical system and probe system.  Incorporating a
nanotube SQUID into a transmon-qubit design would achieve a
single-phonon coupling rate of $g = 7$ MHz (see Supplementary
Information).  Such a coupling strength would be within the
single-phonon strong-coupling limit, providing a means for the readout and
control of mechanical quantum states of a carbon nanotube.

\setlength{\parindent}{0em}

\vspace{2em}

\textbf{Methods} 

Fabrication begins with an oxidized p++ Si wafer (285 nm oxide), in
which the doped substrate is used as a global backgate.  A 40/40 nm
Mo/Re bilayer is deposited by magnetron sputtering, and electrodes
are subsequently defined by reactive ion etching with an SF$_6$
plasma. Reactive ion etching is continued into the substrate, also
defining self-aligned trenches in the SiO$_{2}$.  The Mo and Re in
the two separate layers subsequently alloy together during the
nanotube growth step.  The resulting film is a superconductor with a
$T_{\text{C}}$ of $5.5$ K.  Nanotubes are grown over the structure
in the last step using a methane CVD growth \cite{steele2009tunable},
and promising devices are selected from room-temperature electrical
characterization.  Measurements are performed in a dilution
refrigerator at a base temperature of $25$ mK.  The device is
connected via copper powder filters and low pass filters at base
temperature to the measurement equipment at room temperature.
Measurements are performed in either a 4-terminal current bias or a
2-terminal voltage bias configuration.

\vspace{0em}
\setlength{\parindent}{0em}



\setlength{\parindent}{0pt}

\vspace{1em}

%

\vspace{2em}

{\bf \large Acknowledgments}

This work was supported by the Dutch Organization for Fundamental
Research on Matter (FOM), the Netherlands Organization for Scientific
Research (NWO), and the EU FP7 STREP program (QNEMS).

\vspace{2em}

{\bf \large Author Contributions}

B.H.S. fabricated the sample. B.H.S. and S.E. conducted the
experiments. B.H.S. and G.A.S. wrote the manuscript. G.A.S. and
H.v.d.Z. supervised the project. All authors discussed the results,
analyzed the data, and commented on the manuscript.

\vspace{2em}

\end{document}


\setlength{\parindent}{0em}

{\sf \LARGE Coupling carbon nanotube mechanics to a superconducting
  circuit: Supplementary Information }

\vspace{0.5em}

B.H. Schneider$^1$, S. Etaki$^1$, H.S.J. van der Zant$^1$, G.A. Steele$^1$

\vspace{2em}

{\em $^1$Kavli Institute of NanoScience, Delft University of
Technology, PO Box 5046, 2600 GA, Delft, The Netherlands.}

\vspace{8em}

\section*{S1 Device characterization}

In Fig.\ S1, we present electrical measurements characterizing the
transport properties of the device, in which the Coulomb blockade and
Fabry-Perot transport regimes for different gate votlages can be seen,
and from which the bandgap is estimated.

\begin{figure}[h]
\centering
\includegraphics[width=\linewidth]{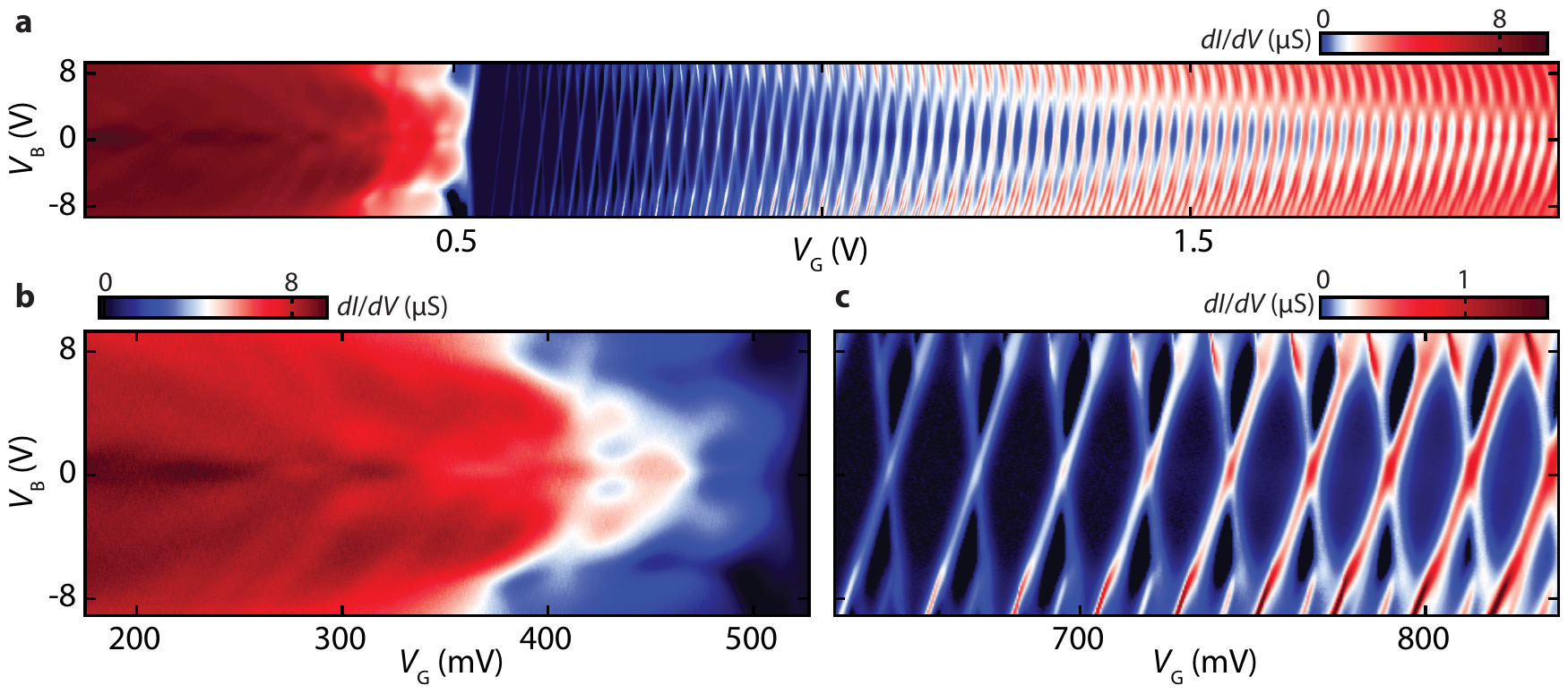}
\caption{ \textbf{Two-terminal differential conductance a,} as a function of
  the source-drain bias voltage \Vb and gate voltage $V_\text{G}$,
  taken at 1.2 K (series resistance from wiring and filters has not
  been subtracted).  This dataset was taken during an earlier cooldown
  of the device in a different cryogenic insert. As a result, there is
  a slight shift of the threshold gate voltage for hole conductance
  compared to Fig.\ 1c of the main text. We determine the bandgap of
  the device from the size of the empty Coulomb diamond by subtracting
  the average of the 1e/1h addition energies from the empty-dot addition
  energy. \textbf{b,} Zoom of the dataset showing the high-conductance
  Fabry-Perot regime when doping the device with holes. \textbf{c,} Zoom of
  dataset showing Coulomb blockade when the device is doped with
  electrons. When doping the device with electrons, tunnel barriers
  naturally form from p-n junctions near the edge of the trench. The
  p-n junctions arise from a gate-independent p-type doping of the
  nanotube near the trench edge due to the work function difference
  between the nanotube and the metal contacts.}
\label{fig:S1}
\end{figure}

\section*{S2 Discriminating gate-induced flux from time-dependent flux creep}

\begin{figure}[th]
\centering
\includegraphics[width=\linewidth]{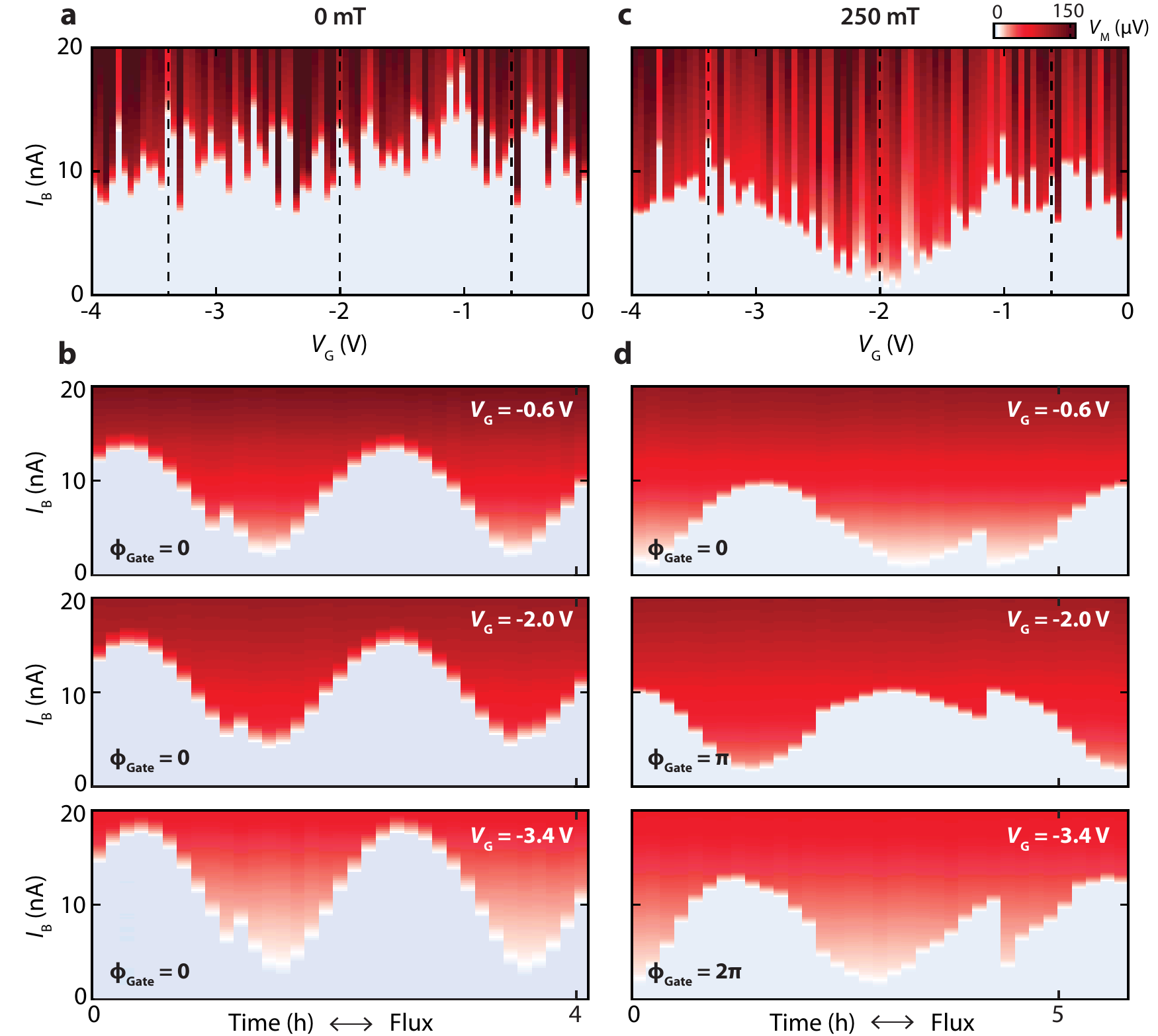}
\caption{ \textbf{Gate-voltage induced flux observed by a gate-dependent
    shift of the phase of the time-dependent oscillations of \Ic a,}
  Voltage \Vm as a function of applied bias current \Ib
  and gate voltage \Vg at $B = 0$. \textbf{b,} \Vm as a function of
  \Ib and of time for three different gate voltages indicated by the
  dashed in \textbf{a}. \textbf{c,d} Same measurement as in \textbf{a,b}, but
  now taken in the presence of a parallel magnetic field of 250
  mT. \textbf{d,} A gate dependent phase shift $\Phi_{Gate}$ of the
  oscillations of $I_\text{C}$ is now observed, demonstrating that the
  gate voltage is inducing a magnetic flux in the SQUID.}
\label{fig:S2}
\end{figure}

In this section, we present measurements which discriminate between
time dependent magnetic flux creep (Fig.\ 2c in the main text) and
gate-induced flux (Fig.\ 3b of the main text). This is done by
plotting the time-dependent oscillations of $I_\text{C}$ at different
gate voltages. If the gate is inducing no flux in the SQUID, the
oscillations at different gate voltages should all be in phase. If the
gate is inducing a flux in the SQUID, there will be a phase shift
between the oscillations at different gate voltages.

\setlength{\parindent}{2em}

In Fig.\ S2, we demonstrate that at $B = 0$, the oscillations in time
at different gate voltages are in phase, while $B = 250$ mT, they are
shifted by the gate-induced flux. The data in Fig.\ S2 are extracted
from a 3-D $(x,y,z) = (I_\text{B},V_\text{G},t)$ dataset (one at $B=0$ and one at
$B=250$ mT).  The measurements are performed by sweeping the bias
current, stepping the gate voltage quickly, and then repeating this in
time. The gate sweep is performed quickly enough such that the
measurement time for a full gate sweep measurement, as shown in
Fig.\ S2a, is fast compared to the external flux drift: the
measurement time for such a gate sweep is $t_{meas} = 7$ min, while
the external flux creep rate during these measurements is about 1
$\Phi_0$ in two hours. Each gate sweep can therefore be considered
to be taken at a fixed external flux. Note that in addition to the
slow flux creep, we also sometimes observe sudden jumps in the
external flux, such as can be seen at $t =$ 4.2 hours in
Fig.\ S2d. This results in a sudden jump in the phase of the
oscillations. The gate induced phase shift, however, can still be
tracked both immediately before and after the jump. The gate traces in
Fig.\ S2 and Fig.\ 3 of the main text are extracted at timesteps where
these flux jumps are not present.

Figure S2b shows $I_\text{C}$ vs.\ time for three different gate voltages at
$B = 0$.  At zero external field, the oscillations of $I_\text{C}$ measured
at different gate voltages are all in phase, indicating no
gate-induced flux. In Fig.\ S2d, we show $I_\text{C}$ oscillations in time at
$B = $250 mT. The d.c.\ gate voltage now shifts the phase of the
oscillations, as can be seen clearly in Fig.\ S2d. This gate-dependent
phase shift demonstrates that the gate-induced sinusoidal modulation of $I_\text{C}$
shown in Fig.\ S2c, and Fig.\ 3c of the main text, are indeed caused
by a gate-voltage induced magnetic flux.

\setlength{\parindent}{0em}

\section*{S3 Expected static displacement of the nanotube with gate voltage}

\begin{figure}[th]
\centering
\includegraphics[width=3in]{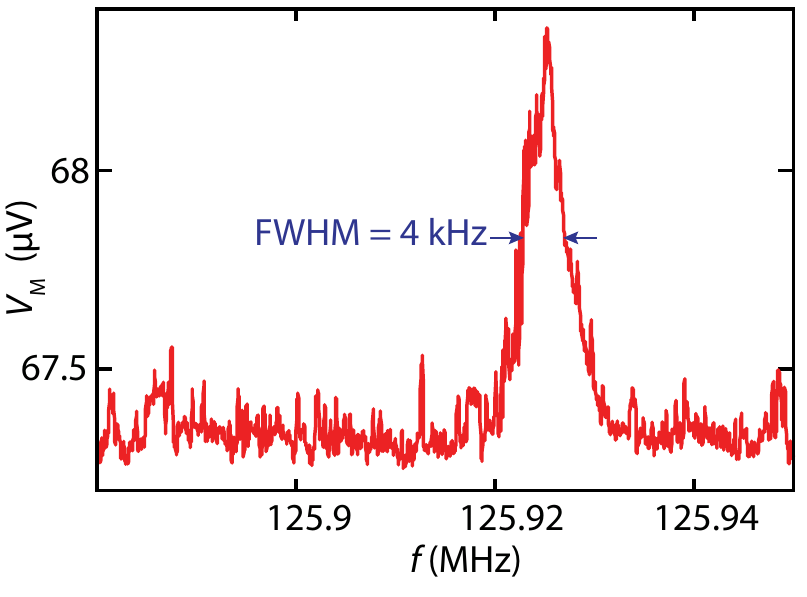}
\caption{\textbf{Driven mechanical resonance of the nanotube} measured by
  applying a bias current above the critical current and using
  rectification readout technique introduced in
   \cite{huettel2009carbon}. We estimate the quality factor of the
  resonance from the full width at half maximum, $\Delta f$, of the
  measured curve. With $f_{\text{R}}=126$ MHz and $\Delta f =$ 4 kHz,
  we get $Q = f_{\text{R}}/ \Delta f = 3\times 10^4$.}
  \label{fig:S3}
\end{figure}

When a constant voltage $V_{\text{G}}$ is applied to the gate, the
suspended nanotube segments are attracted to the gate by a Coulomb
force, $F_{\text{C}} \propto V_{\text{G}}^2$. The equilibrium position
of the nanotube corresponds to the position where this Coulomb force
is balanced by the mechanical restoring force
 \cite{sapmaz2003carbon}. At small gate voltages, the bending rigidity
of the nanotube dominates the mechanical restoring force, giving a
static displacement $u\propto V_{\text{G}}^2$ (the weak bending
regime). Beyond a certain gate voltage, induced tension in the
nanotube becomes important in determining the mechanical restoring
force, and there is a transition to a strong bending regime in which
$u\propto V_{\text{G}}^{2/3}$. The transition between these two
regimes depends on the dimensions of the nanotube, and can also be
influenced by additional tension introduced, for example, by the
fabrication process. In any case, the net result is that the
displacement of the nanotube as a function of gate voltage is, to a
good approximation, linear over a relatively wide regime of voltages,
as can be seen in Fig.\ 2 of Sapmaz {\em et al.}
 \cite{sapmaz2003carbon}.

\setlength{\parindent}{2em}

If the nanotube displacement was not linear in gate voltage, the
periodicity of the $I_\text{C}$ oscillations in gate voltage, $\Delta
V_G^{-1}$, would change slowly as a function of gate voltage. The
relatively constant $\Delta V_G^{-1}$ we observe in Fig. 4a of the
main text indicates that the nanotube displacement in our device is
indeed approximately linear in the range of gate voltages we study.

Critical current oscillations in gate voltage were fitted to a cosine function:
\begin{equation}
f(x) = \frac{a+b}{2} + \frac{a-b}{2}\cos\left[(x-x_{0})\frac{2\pi}{L}\right]
\label{eq:fitting}
\end{equation}
where $a$ and $b$ are the maximum and minimum of the modulation respectively, and where $x_{0}$ is the position at the maximum and $L$ the periodicity.

\setlength{\parindent}{0em}

\section*{S4 Definition of the nanotube displacement and estimation
 of the flux responsivity}

In this section, we give a rigorous definition of the nanotube
displacement, and use this definition to calculate the flux
responsivity of the device. In particular, following Poot {\em et al.} \cite{poot} ,
we define the displacement of a mode of the nanotube in such a way that
we require only one effective mass for all modes, avoiding the
complication of having different effective masses for different modes.

\setlength{\parindent}{2em}

The zero-frequency, or dc, flexural displacement $z_{\text{dc}}(x,y)$
of the carbon nanotubes towards the back gate (the $x-y$ plane), can
be described by a single coordinate $u_{\text{dc}}$. The displacement
per unit force and the change of area $\Delta A$ per unit displacement
both depend on the chosen definition of $u_{\text{dc}}$. This is also
true for the mechanical resonance modes of the nanotube, which form an
eigenbasis for the nanotube displacement. Any periodic displacement of
the nanotube with frequency $f$ can be decomposed into a superposition
of the eigenmodes:
\begin{equation}
z(x,y,t)=\sum_0^{\infty}u_n\xi_n(x,y)\cos{(2\pi f t+\varphi_n)},
\end{equation}
where $u_n$ is the displacement coordinate, $\xi_n(x,y)$ is the
normalized mode shape, $f_n$ is the eigenfrequency and $\varphi_n$
is the phase offset of mode $n$. For a nanotube with length $\ell$
much larger than its cross-sectional diameter, $\xi_n(x,y)$ is
usually integrated in the radial direction, such that the mode shape
can be described as a function of only the distance along its length
direction, $x$, i.e. $\xi_n(x,y)\rightarrow\xi_n(x)$. The dc
displacement ($f=0$) is related to the eigenmodes by:
\begin{equation}
z_{\text{dc}}(x)=\sum_0^{\infty}u_n\xi_n(x).
\end{equation}
In general, the displacement, modeshapes and eigenfrequencies of
the nanotube must be solved from its elastodynamic differential
equations and depend on the nanotube geometry, its rigidity, any
built-in tension, and the amount and distribution of applied forces.
Once this is done, the definition of displacement depends on
the choice of normalization for $\xi_n(x)$. A convenient
normalization for $\xi_n(x)$ is:
\begin{equation}
\frac{1}{\ell}\int_0^{\ell}\xi_n(x)^2dx=1.
\end{equation}
With this normalization, the displacement coordinate $u_n$ is
(spatial) root-mean-square displacement of mode $n$. The dynamical
spring constant of each eigenmode now equals $k_n=m_{\text{R}}(2\pi
f_n)^2$, where $m_{\text{R}}$ is the nanotube mass. The
change in area due to a d.c.\ nanotube displacement becomes:
\begin{align}
\Delta A=&\int_0^{\ell}z_{\text{dc}}(x)dx=\sum_0^{\infty}a_n\ell u_n,\\
a_n\equiv&\frac{1}{\ell}\int_0^{\ell}\xi_n(x)dx.
\end{align}
To estimate the numerical coefficients $a_n$, we assume sinusoidal
eigenmode shapes for the nanotubes: $\xi_n(x)=\sqrt{2}\sin{(\pi
nx/\ell)}$ (based on  \cite{poot}). The numerical coefficients
then become $a_n=0.9/n$ for odd $n$ and $a_n=0$ for even $n$ (no net
area change). For a displacement due to a uniformly distributed dc
force, the amplitude of each eigenmode is proportional to
$(a_n/f_n)^2$, which means that the shape of the dc deflection is
almost entirely determined by the fundamental eigenmode ($f_n$ is
roughly proportional to $n$). The dc spring constant is then equal
to $k_1$ and the area change for both $u_{\text{dc}}$ and $u_1$ is
characterized by the same coefficient, $a_1=0.9$.

Having defined the displacement, we are now in a position to calculate
the responsitivity of the device. The flux responsivity $\Phi_u$ which
we give in the main text is calculated by multiplying $\Delta A$ with
the applied magnetic field $B$ and dividing out the displacement
$u_{\text{dc}}$:
\begin{equation}
\Phi_u\equiv\frac{d\Phi}{du_{\text{dc}}}\approx\frac{d\Phi}{du_1}=a_1B\ell.
\end{equation}
At a field of 1 T and a suspended nanotube length of 800 nm, we get a
responsivity $\Phi_u=0.35$ $\Phi_0$/nm per suspended nanotube
segment. The dc displacement of the nanotube due to the applied gate
voltage can now be calculated based on Fig. 4a of the main text: At 1
T, we observe five oscillations of the SQUID critical current,
i.e. $\Delta \Phi=5\Phi_0$. With the calculated responsivity, the
displacement of each nanotube segment over the full gate voltage range
is $\Delta\Phi/\Phi_u=7$ nm.

\setlength{\parindent}{0em}

\section*{S4 Estimation of the zero point motion}

Quantum mechanical displacement fluctuations are dominant when a
mechanical resonator with resonance frequency $f_{\text{R}}$ is cooled
to a temperature $T$ such that its thermal energy is far less than the
energy of a single phonon, i.e. $k_{\text{B}}T\ll hf_{\text{R}}$.
Here, $k_{\text{B}}$ is the Boltzmann constant and $h$ is the Planck
constant. In this regime, the resonator has an average phonon
occupation which approaches zero, and displacement
fluctuations are due to the quantum mechanical ground state energy of
the resonator, which equals that of half a phonon. The
root-mean-square value of the ground state displacement fluctuations
is called the zero-point motion and is given by  \cite{poot}:
\begin{equation}
u_{\text{zpf}}=\sqrt{\frac{hf_{\text{R}}}{2m_{\text{R}}(2\pi
f_{\text{R}})^2}}
\end{equation}
The maximum power spectral density $S_{uu}(f)$
due to the zero-point motion occurs at the resonance frequency and
is related to $u_{\text{zpf}}$ according to  \cite{poot}:
\begin{equation}
S_{uu}^{\text{zpf}}(f_{\text{R}})=u_{\text{zpf}}^2\left(\frac{\pi
f_{\text{R}}}{2Q}\right)^{-1}=\frac{hQ}{\pi
m_{\text{R}}(2\pi f_{\text{R}})^2}
\end{equation}
where $Q$ is the quality factor of the resonator. In order to measure
the zero-point fluctuations, the measurement sensitivity of the
detector should be better (lower) than
$S_{uu}^{\text{zpf}}(f_{\text{R}})$. The suspended carbon nanotubes in
this paper each have a fundamental eigenmode with displacement in the
direction of the back gate. Figure S3 shows a measurement of the
mechanical response of the fundamental mode of one of the nanotubes to
an applied driving force. From the response curve, we find a resonance
frequency of 126 MHz and a quality factor $Q = 3\times 10^4$. The mass
of an 800 nm long single-walled carbon nanotube is approximately
$m_{\text{R}}=5\times 10^{-21}$ kg (and the corresponding spring constant
is thus $k=3\times10^{-3}$ N/m). Using this lower bound for $Q$ gives
$u_{\text{zpf}}=3.6$ pm and
$\sqrt{S_{uu}^{\text{zpf}}(f_{\text{R}})}=45$
fm/$\sqrt{\text{Hz}}$. With the above responsivity, the zero-point
fluctuations of a single suspended nanotube segment result in a flux
noise in the SQUID of 16 $\mu \Phi_0 / \sqrt{\text{Hz}}$.

\section*{S5 Estimates of coupling for a nanotube transmon qubit}

The zero-phonon coupling rate $g$ is given by the shift in the energy
levels of the qubit in response to the zero-point fluctuations of the
nanotube position \cite{lahaye2009nanomechanical,painter}. The zero
point fluctuations of $u_{\text{zpf}}=3.6$ pm together with the
responsivity of 0.35 m$\Phi_0$/pm gives a correponding flux shift of
$\Phi_{ZPF} = 1.3$ m$\Phi_0$. In the transmon
limit, where the charging energy is much smaller
than the Josephson energy ($E_C \ll E_J$), the energy splitting of the
qubit is given by \cite{koch2007charge}:
\begin{equation}
E_{01} \approx \sqrt{8 E_J E_C}
\end{equation}
A small change in the qubit energy due to a change in the Josephson
energy is then given by:
\begin{equation}
\delta E_{01} = \frac{E_{01}}{2} \frac{\delta E_J}{E_J}
\end{equation}
We now need to estimage $\delta E_J$ in response to the $\Phi_{ZPF}$
above. Our SQUID shows a near complete suppression of the critical
current as a function of flux, allowing us to estimate the change in
$E_J$ on the slope of the flux oscillation as
\begin{equation}
\frac{dE_J}{d\Phi} \approx \frac{E_J^{max}}{0.5\Phi_0}
\end{equation}
The coupling rate $g$, given by the shift of the qubit energy in response to the zero point
fluctuations of the nanotube position, can then be estimated as:
\begin{equation}
g = \delta E_{01}^{ZPF} \approx \frac{E_{01}}{2} \frac{\Phi_{ZPF}}{0.5\Phi_0}
\end{equation}
Assuming $E_{01}$ is designed to be 6 GHz, and using $\Phi_{ZPF} =
1.3$ m$\Phi_0$, we estimate the coupling to be $g = 7$ MHz.

\vspace{0em}
\setlength{\parindent}{0em}